# Dissipative and quantum mechanics

Roumen Tsekov
DWI, RWTH, 52056 Aachen, Germany

Three existing interpretations of quantum mechanics, given by Heisenberg, Bohm and Madelung, are examined to describe dissipative quantum systems as well. It is found that the Madelung quantum hydrodynamics is the only correct approach. A new stochastic reinterpretation of the quantum mechanics is proposed, which represents the microscopic face of the Madelung hydrodynamics. The main idea is that the vacuum fluctuates permanently, which explains the probabilistic character of the quantum mechanics. Thus, it is an objective theory independent of the human beings and their measurements. The effect of the thermal fluctuations in the surrounding is also accounted for via a heuristic Langevin equation with two random forces. Some statistical characteristics of these quantum and thermal noises are determined by reproducing known results for the system phase-space dynamics.

The quantum mechanics is formulated for closed systems possessing Hamiltonians. From the classical mechanics it is known, however, that dissipative systems are not conservative and cannot be described via the canonical formalism. In fact, Caldirola [3] and Kanai [9] have proposed a time-dependent Hamiltonian to quantize dissipative systems but the momentum in their model differs from the classical expression. As a result, the Caldirola-Kanai model violates the Heisenberg uncertainty principle. There are also other attempts to formulate a dissipative quantum mechanics (for review see [6, 19, 26]). From the quantum mechanical point of view the best solution is a reduction of the Hamiltonian dynamics of a closed system to those of a subsystem and environment. However, there is still no satisfactorily explanation how the irreversibility can originate from a reversible dynamics and, perhaps, the best solution of this so-called Loschmidt paradox is given by the chaos concept. Practically, all many-particle systems are non-integrable and these so-called large Poincare systems are described better in terms of stochastic trajectories rather than by a canonical dynamics [15, 16], since any employed Hamiltonian is approximate to the real one. Also, due to infinite range of the fundamental interactions, there are no isolated systems in the Nature at all. Of course, when subsystem under consideration is far from others, the effect of the surroundings is very weak but if the subsystem is chaotic this influence could be strong enough to make the whole dynamics unpredictable.

A way to avoid the use of any Hamiltonian is just to extend the dissipative classical Newtonian equation of motion for a particle with coordinate $R$

$$m\ddot{R} + b\dot{R} = -\nabla U \qquad (1)$$

to a quantum equation of motion via the Heisenberg time-dependent operator formalism. Here $b$ is the particle friction coefficient and $U$ is an external potential. Using the operator version of Eq. (1), one can study, for instance, what happens with the fundamental commutator. It can be easily shown that $[\hat{r}(t), \hat{p}(t)] = i\hbar \mathbb{I} \exp(-bt/m)$, where the initial commutator is taken as the standard value $i\hbar \mathbb{I}$ with $\mathbb{I}$ being the unit tensor. This expression means that the particle loses its quantum character and after sufficiently large time ($t \gg m/b$) it becomes a classical one with commuting position and momentum. Obviously, the Heisenberg matrix mechanics is also not able to describe properly dissipative quantum systems. This is not surprising since the definition of the Heisenberg time-dependent operators requires a system Hamiltonian. The Hamiltonian mechanics, however, is a subclass of the more general Newtonian mechanics, a particular example of which is Eq. (1).

Another way to describe dissipative quantum systems is the Bohmian mechanics [1]. One can extend the de Broglie-Bohm pilot-wave description of a conservative quantum system to a dissipative one by the following dissipative quantum Newtonian equation

$$m\ddot{R} + b\dot{R} = -\nabla U - \nabla Q \qquad (2)$$

As is seen, the quantization in the Bohmian mechanics simply adds to Eq. (1) the quantum potential $Q \equiv -\hbar^2 \nabla^2 \sqrt{\rho} / 2m\sqrt{\rho}$, which depends on the probability density $\rho(r,t)$ to find the particle at a given point $r$ at time $t$. Thus, the relation between the quantum and classical mechanics is clear: if the mass $m$ of the particle is large enough that the quantum force becomes negligible in Eq. (2), the particle behaves as a classical one. Introducing now the action $S$ via the standard relation from the classical mechanics $\dot{R} = \nabla S(R,t)/m$, Eq. (2) can be transformed after integration into a dissipative Bohmian Hamilton-Jacobi equation

$$\partial_t S + (\nabla S)^2 / 2m + U + Q = -bS/m \qquad (3)$$

The frictional term on the right hand side leads to loss of action in time. Since the quantum potential depends on the probability density $\rho$, Eq. (3) is bounded to the continuity equation [1]

$$\partial_t \rho + \nabla \cdot (\rho \nabla S / m) = 0 \qquad (4)$$

Introducing the Bohm presentation of the wave function $\psi = \sqrt{\rho} \exp(iS/\hbar)$, the coupled system of Eqs. (3) and (4) can be transformed into a dissipative Schrödinger equation

$$i\hbar \partial_t \psi = [-\hbar^2 \nabla^2 / 2m + U - i\hbar b \ln(\psi/\psi^*)/2m]\psi \qquad (5)$$

The frictional term in Eq. (5), proposed first by Kostin [10] in his Schrödinger-Langevin equation, is nonlinear and, hence, the superposition principle is not valid for dissipative quantum systems. Thus, the dissipation destroys completely the linear quantum Hilbert space and, for this reason, it could not be described by the Heisenberg formalism as was shown in the beginning. Therefore, for dissipative systems the Schrödinger and Heisenberg pictures are not only different but, in fact, there is no Heisenberg picture at all for the case of nonlinear Schrödinger equations.

In the strong friction limit the action $S$ is small due to the slow motion and in this case one can neglect the first two inertial terms in Eq. (3) to obtain $S = -m(U+Q)/b$. Introducing now this expression in Eq. (4) yields a purely quantum diffusion equation [21]

$$\partial_t \rho = \nabla \cdot [\rho \nabla (U+Q)/b] \quad (6)$$

It is clear that the quantum potential is, in fact, a non-thermal chemical potential of a particle in vacuum, which is driving the quantum diffusion. This is also evident from the fact that the mean value of $Q$ is proportional to the Fisher entropy [17]. In the case of a harmonic oscillator with $U = m\omega_0^2 r^2/2$, where $\omega_0$ is the oscillator own frequency, the solution of Eq. (6) is a Gaussian distribution density with the following position dispersion [21]

$$\sigma^2 = (\hbar/2m\omega_0)\sqrt{1-\exp(-4m\omega_0^2 t/b)} \quad (7)$$

In accordance to the Heisenberg principle the oscillator possesses initially infinite kinetic energy, which is dissipated in time to get finally at the ground state with position dispersion $\hbar/2m\omega_0$. In the case of a free diffusing quantum particle ($\omega_0 = 0$) the already known expression [20] $\sigma^2 = \hbar\sqrt{t/mb}$ follows from Eq. (7).

Using Eq. (7) one can calculate the action value $S(t \to \infty) = -3m\hbar\omega_0/2b$ at equilibrium, which appears to be constant. Hence, according to Bohm [1] the particle will not move in the ground state since $\dot{R} = \nabla S/m = 0$. This contradicts, however, to the quantum mechanics and shows inconsistency of the Bohmian mechanics. According to Eq. (2) the Bohmian particle obeys a deterministic trajectory, which disagrees with the probabilistic nature of the quantum mechanics. The reason for this problem is the de Broglie-Bohm interpretation of the wave function as a physical field guiding the particle, which on the other hand provides the probability density of finding the particle at a given place. This philosophical discrepancy can be resolved by replacing of the de Broglie-Bohm guiding equation $\dot{R} = \nabla S/m$ via $V = \nabla S/m$ [11, 24]. Now, $V(r,t)$ is the velocity of the particle passing through the point $r$ but averaged along all its realizations and $S$ is its hydrodynamic potential. Hence, the condition $\nabla S = 0$ does not mean that the particle is in rest ($\dot{R} = 0$) but that at different experiments the quantum particle will pos-sess different velocities, which will cancel each other in average to obtain $V=0$. Using the Madelung transformation [11] $\psi = \sqrt{\rho}\exp(im\int V \cdot dr/\hbar)$, Eq. (5) reduces to the following two equations

$$\partial_t \rho + \nabla \cdot (\rho V) = 0 \qquad\qquad m\partial_t V + mV\cdot\nabla V + bV = -\nabla(U+Q) \qquad (8)$$

which are analogical to Eqs. (4) and (3), respectively. However, even if the Madelung hydrodynamics is usually considered as a precursor of the Bohmian mechanics, the philosophies behind the de Broglie-Bohm and Madelung theories are completely different. Equations (8) describe quantum diffusion without any hidden variables. This statistical interpretation of the quantum mechanics is more reasonable than the de Broglie-Bohm theory, since the latter is an inconsistent mix of micro- and macroscopic descriptions. That is why the Bohmian mechanics contradicts to the quantum mechanics at some points such as in static stationary states [1], etc. [13]. For instance, there are real wave functions as solutions of the stationary Schrödinger equation. Hence, according the Bohmian mechanics the particles are not moving but on the other hand they possess average kinetic energy according to quantum mechanics. The reason is that $Q$ is not potential but kinetic energy stored as the Fisher entropy. The Madelung hydrodynamics is not, in fact, a hydrodynamics at all, since Eqs. (8) describe the motion of a single quantum particle. The confusion is coming from the similarity between the hydrodynamic and probabilistic equations even in the case of many particles [24]. The frictional term in Eq. (8b) is not, however, hydrodynamic, since it is not proportional to $\nabla^2 V$ [25], which is typical for a Newtonian viscous fluid.

How it was already mentioned $Q$ is a non-thermal chemical potential. In the case when the classical surrounding experiences thermal motion as well, the total chemical potential is expected to be given by $\mu = \mu_0 + k_B T \ln\rho + Q$ [23], where $T$ is the temperature. As is known, the logarithmic term here originates from the Boltzmann-Shannon entropy, while the quantum potential corresponds to the local Fisher entropy. In the high friction limit the thermodynamically generalized equation (8b) reads $V = -\nabla(U+\mu)/b$ [23]. Introducing this expression for $V$ in the continuity equation (8a) results in the following thermo-quantum diffusion equation [21]

$$\partial_t \rho = \nabla\cdot[\rho\nabla(U+Q)/b + D\nabla\rho] \qquad (9)$$

where $D = k_B T/b$ is the Einstein diffusion constant. The solution of Eq. (9) for a free Brownian particle ($U=0$) is a Gaussian density with position dispersion given by the relation [20, 21]

$$\sigma^2 - \lambda_T^2 \ln(1+\sigma^2/\lambda_T^2) = 2Dt \qquad (10)$$

where $\lambda_T = \hbar/2\sqrt{mk_B T}$ is the thermal de Broglie wave length. This expression is a quantum generalization of the classical Einstein law $\sigma^2 = 2Dt$, which follows from Eq. (10) at $\hbar \to 0$. In the case of zero temperature Eq. (10) reduces to the already mentioned expression $\sigma^2 = \hbar\sqrt{t/mb}$ for the purely quantum diffusion. The thermo-quantum description above corresponds to the following Schrödinger equation [5, 23]

$$i\hbar\partial_t \psi = [-\hbar^2\nabla^2/2m + U + k_B T \ln(\psi\psi^*) - i\hbar b \ln(\psi/\psi^*)/2m]\psi \qquad (11)$$

As is seen, the Boltzmann entropy leads to another logarithmic term in Eq. (11). Hence, the irreversible thermodynamics makes the Schrödinger equation nonlinear but, on the other hand, the quantum effects make nonlinear the diffusion equation (9) as well. Due to the second and third laws of thermodynamics, respectively, there are no systems without friction and the absolute zero temperature is impossible to reach. Therefore, the nonlinear terms in Eq. (11) could never vanish for real physical systems. This questions the reality of many quantum effects following from the superposition principle.

In contrast to the Bohmian mechanics, the Madelung quantum hydrodynamics provides only statistical information about the real motion of the particle, which is obviously stochastic. Hence, an open question is what stochastic picture is hidden behind the Madelung hydrodynamics. How it was already discussed, Eq. (2) is an improper mix of mechanical and statistical descriptions [24]. Thus, in a correct model the macroscopic Bohmian quantum force $-\nabla Q$ should be replaced by a microscopic stochastic quantum force $f_Q$ to obtain

$$m\ddot{R} + b\dot{R} = -\nabla U + f_Q \qquad (12)$$

This stochastic equation should not be confused by the Langevin equation, because the random force $f_Q$ originates purely from vacuum fluctuations and it is present even for a single particle in vacuum ($b=0$). This random force possesses a zero mean value to satisfy the Ehrenfest theorem. Equation (12) represents a new stochastic reinterpretation of the quantum mechanics, according to which the vacuum fluctuates permanently and for this reason the trajectory of a particle in vacuum is random. If the particle is, however, too heavy the vacuum fluctuations generate negligible forces and this particle obeys the laws of classical mechanics. That is why the quantum mechanics is important for light particles. The probabilistic character of the quantum mechanics originates now from the persistent vacuum fluctuations and is not related to any measurements, how it is assumed in the Copenhagen interpretation. The Planck constant $\hbar$ is the main characteristic of the vacuum fluctuations. The force $f_Q$ in Eq. (12) explains naturally the so-called 'tunnel effect' without any tunneling, in a why the thermal fluctuations assist barrier overcoming. It produces also the zero-point quantum energy. Finally, the non-locality of quantum mechanics could be understood via the spatial correlations of the quantum stochastic force. The present causal theory is a stochastic enhancement of the de Broglie-Bohm theory, where the vacuum fluctuations are guiding randomly the particles. It should not be mixed, however, by the Bohm-Vigier model [2] introducing some macroscopic fluctuations in the Madelung fluid. An alternative of the present model is the Fenyes-Nelson stochastic theory [7, 12], which presumes also ad hoc universal vacuum fluctuations, taking place forward and backward in time. The drift terms in the Nelson stochastic equations, however, depend on the probability density, an indication of a macroscopic description similar to the Bohm-Vigier model.

Let us now try to reproduce the Madelung hydrodynamics from Eq. (12). The probability density is given by $\rho = <\delta(r-R)>$, where the brackets $<>$ denote statistical average over the realizations of $R$. Taking a time derivative of this presentation and comparing the result with the continuity equation (8a) yields the following standard expression $V = <\dot{R}\delta(r-R)>/\rho$ for the hydrodynamic velocity. It is clear now that $V=0$ does not mean $\dot{R}=0$ unless the particle motion is not random (mathematically equivalent to removal of the brackets $<>$), how it is in the classical mechanics. This is the origin of the inconsistency of the Bohmian mechanics, where the particle motion is considered deterministic with $V=\dot{R}$. Taking a time derivative of $V$ and employing Eq. (12) one yields the following hydrodynamic momentum balance

$$m\partial_t V + mV\cdot\nabla V + bV = -\nabla U - [\nabla\cdot\mathbb{P} - <f_Q\delta(r-R)>]/\rho \qquad (13)$$

where $\mathbb{P} = <m(\dot{R}-V)(\dot{R}-V)\delta(r-R)>$ is the pressure tensor. Due to the vacuum isotropy the random quantum force is not correlated with the position of the particle. In this case the last term in Eq. (13) omits since $<f_Q\delta(r-R)> = <f_Q>\rho = 0$. The force $f_Q$ still determines the particle motion via the pressure tensor. This is evident from the fact that if the random force is canceled than $\dot{R}=V$ and hence $\mathbb{P}=0$. Comparing now Eqs. (13) and (8b) leads to $\nabla\cdot\mathbb{P}=\rho\nabla Q$, which confirms the interpretation of $Q$ as a quantum chemical potential. Integrating this relation yields already known expression for the quantum pressure tensor $\mathbb{P} = -(\hbar^2/4m)\rho\nabla\nabla\ln\rho$ [18].

How it was shown in the case of thermo-quantum diffusion the chemical potential equals to $\mu = \mu_0 + k_BT\ln\rho + Q$, which according to the thermodynamic Gibbs-Duhem isotherm $\nabla\cdot\mathbb{P} = \rho\nabla\mu$ corresponds to $\mathbb{P} = k_BT\rho\mathbb{I} - (\hbar^2/4m)\rho\nabla\nabla\ln\rho$ [21]. The first term here represents the ideal gas thermal pressure. The microscopic picture behind these macroscopic expressions is given by the following stochastic equation

$$m\ddot{R} + b\dot{R} = -\nabla U + f_Q + f_L \qquad (14)$$

where $f_L$ is the Langevin force accounting for the thermal fluctuations in the surrounding. One can define generally the phase-space distribution function via $W = <\delta(p-m\dot{R})\delta(r-R)>$. Taking a time-derivative of $W$ and using Eq. (14) yields a thermo-quantum diffusion equation

$$\partial_t W + \frac{p}{m} \cdot \nabla W - \nabla U \cdot \partial_p W = \partial_p \cdot [b\frac{p}{m}W - <(f_Q + f_L)\delta(p-m\dot{R})\delta(r-R)>] \quad (15)$$

In the classical limit the quantum force vanishes, while the Langevin force term acquires the form $<f_L \delta(p-m\dot{R})\delta(r-R)> = -\int <f_L(t)f_L(s)> \cdot \partial_p W(p,r,s) ds = -bk_B T \partial_p W$. The second expression here is due to the Furutsu-Novikov theorem [8, 14] for Gaussian processes, while the last result follows from the autocorrelation function $<f_L(t)f_L(s)> = 2bk_B T \delta(t-s)\mathbb{I}$ of the classical Langevin force. Thus, Eq. (15) reduces to the classical Klein-Kramers equation

$$\partial_t W + \frac{p}{m} \cdot \nabla W - \nabla U \cdot \partial_p W = b\partial_p \cdot (\frac{p}{m}W + k_B T \partial_p W) \quad (16)$$

In the case of a particle in vacuum $b=0$ and the Langevin force vanishes. The motion of the quantum particle in the phase-space is described than via the Wigner-Liouville equation [27]

$$\partial_t W + \frac{p}{m} \cdot \nabla W - \sum_{k=0}^{\infty} \frac{(\hbar/2i)^{2k}}{(2k+1)!} \nabla^{2k+1} U \cdot \partial_p^{2k+1} W = 0 \quad (17)$$

which is the phase-space representation of the Schrödinger equation. Note that Eq. (17) cannot be derived in the frames of the Bohmian mechanics. Comparing this equation with Eq. (15) yields the following expression for the quantum force term

$$<f_Q \delta(p-m\dot{R})\delta(r-R)> = -\sum_{k=1}^{\infty} \frac{(\hbar/2i)^{2k}}{(2k+1)!} \nabla^{2k+1} U \cdot \partial_p^{2k} W \quad (18)$$

which diminishes at $\hbar \to 0$. Because the right hand side of Eq. (18) depends only on the current value of $W$ it seems that the quantum noise is also not correlated at different time moments. Obviously, the Furutsu-Novikov theorem does not apply to the quantum fluctuations, since they are not Gaussian. According to Eq. (18) the vacuum fluctuations are sensitive to the external potential, which explains in particular the double-slit experiment. This is a manifestation of the spatial correlations of the vacuum fluctuations reflecting in the quantum non-locality [24]. In contrast to the Langevin force, the quantum random force does no work in average since

$< f_Q \dot{R} >= 0$ according to Eq. (18). It is also interesting that in the cases of a free particle and a harmonic oscillator $f_Q$ is not correlated either to the position or to the momentum of the particle since $< f_Q \delta(p - m\dot{R})\delta(r - R) >= 0$. In the general case one can propose the following expression

$$< (f_Q + f_L)\delta(p - m\dot{R})\delta(r - R) >= -\sum_{k=1}^{\infty} \frac{(\hbar/2i)^{2k}}{(2k+1)!} \nabla^{2k+1} U \cdot \partial_p^{2k} W - bk_B T \partial_p W - b\hat{X}W \qquad (19)$$

where the thermo-quantum operator $\hat{X}$ is still unknown. Introducing Eq. (19) in Eq. (15) leads to a Wigner-Klein-Kramers equation

$$\partial_t W + \frac{p}{m} \cdot \nabla W - \sum_{k=0}^{\infty} \frac{(\hbar/2i)^{2k}}{(2k+1)!} \nabla^{2k+1} U \cdot \partial_p^{2k+1} W = b\partial_p \cdot (\frac{p}{m} W + k_B T \partial_p W + \hat{X}W) \qquad (20)$$

Coffey at al. [4] have obtained a semiclassical expression for $\hat{X} = (\hbar^2 / 12mk_B T)\nabla\nabla U \cdot \partial_p$ on the base of the known equilibrium distribution from the statistical thermodynamics [27]. A possible model leading to the thermo-quantum pressure tensor $\mathbb{P} = k_B T \rho \mathbb{I} - (\hbar^2 / 4m)\rho \nabla\nabla \ln \rho$ is

$$\hat{X} = -(\hbar^2 / 4m)\nabla\nabla \ln \rho \cdot \partial_p \qquad (21)$$

Similar nonlinear expression was derived via generalization of the semiclassical approach of Coffey at al. [22]. For a Gaussian density $\rho$ Eq. (21) acquires the form $\hat{X} = (\hbar^2 / 4m\sigma^2)\partial_p$, which is similar to the thermal term with a quantum temperature [18] given by the minimal Heisenberg momentum uncertainty. Equation (21) is, however, approximate, since Eq. (20) accomplished by Eq. (21) does not provide the exact equilibrium distribution. This is also the case of the corresponding Eq. (9), which is an approximation of the following nonlinear quantum Smoluchowski equation [20]

$$\partial_t \rho = D\nabla \cdot [\rho\nabla\int_0^\beta (U+Q)_b d\beta + \nabla\rho] \qquad (22)$$

The integral on the reciprocal temperature $\beta = 1/k_B T$ reflects the temperature dependence of the quantum potential, which means that, in general, there are additional thermo-quantum terms in the pressure tensor $\mathbb{P}$ and the chemical potential $\mu$ as well. For instance, the chemical potential corresponding to Eq. (22) is $\mu = \mu_0 + k_B T \ln\rho + k_B T \int_0^\beta Q d\beta$. The numerical solution of Eq. (22) for the case of a free quantum Brownian particle is given in [22].

## Appendix

The Appendix aims to explore the Madelung quantum hydrodynamics in momentum space of a quantum harmonic oscillator with own frequency $\omega_0$. Fourier transformation of the relevant equation (1) yields the following Schrödinger equation

$$i\hbar \partial_t \phi = (p^2/2m - m\omega_0^2 \hbar^2 \nabla_p^2 / 2)\phi \tag{23}$$

where $\phi(p,t)$ is the wave function in the momentum $p$-representation. Using the Madelung transformation $\phi = \sqrt{\rho}\exp(iS/\hbar)$, Eq. (23) splits into two real equations

$$\partial_t \rho = -\nabla_p \cdot (\rho m\omega_0^2 \nabla_p S) \tag{24}$$

$$\partial_t S + p^2/2m + m\omega_0^2 (\nabla_p S)^2 / 2 + Q_p = 0 \tag{25}$$

The first one in the continuity equation of the probability density $\rho(p,t)$ in the momentum space, while the second equation has the structure of the quantum Hamilton-Jacobi equation (3). The Bohm quantum potential for a harmonic oscillator in the momentum space is defined via the expression $Q_p \equiv -m\omega_0^2 \hbar^2 \nabla_p^2 \sqrt{\rho} / 2\sqrt{\rho}$ [28]. In accordance to Eq. (11), the dissipative extension of Eq. (25) at arbitrary temperature $T$ reads

$$\partial_t S + p^2/2m + m\omega_0^2 (\nabla_p S)^2 / 2 + Q_p + k_B T \ln\rho = -bS/m \tag{26}$$

At strong friction one can neglect the small $S$-terms and express the action $S$ in the form

$$S = -(p^2/2 + mQ_p + mk_B T \ln\rho)/b \tag{27}$$

Introducing now Eq. (27) in Eq. (24) results in a quantum Fokker-Plank equation

$$\partial_t \rho = (m\omega_0^2 / b)\nabla_p \cdot [p\rho - (m\hbar\omega_0 / 2)^2 \nabla_p (\rho \nabla_p^2 \ln\rho) + mk_B T \nabla_p \rho] \tag{28}$$

The solution of Eq. (28) is the Gaussian distribution $\rho = \exp(-p^2/2\sigma_p^2)/\sqrt{2\pi\sigma_p^2}$ with momentum dispersion satisfying the following equation

$$\partial_t \sigma_p^2 = -2(m\omega_0^2/b)[\sigma_p^2 - (m\hbar\omega_0/2)^2/\sigma_p^2 - mk_BT] \qquad (29)$$

The exact solution of Eq. (29) is quite complicated but some limiting cases could point the correctness of our description. For the classical oscillator Eq. (29) provides the well-known expression $\sigma_p^2 = mk_BT[1-\exp(-2m\omega_0^2 t/b)]$. Another interesting case is at zero temperature, where the solution of Eq. (29)

$$\sigma_p^2 = (m\hbar\omega_0/2)\sqrt{1-\exp(-4m\omega_0^2 t/b)} \qquad (30)$$

corresponds well to our previous results for the position dispersion of the quantum oscillator [21], see also Eq. (7). Finally, at equilibrium the momentum dispersion equals to

$$\sigma_p^2 = (m/2)[\sqrt{(k_BT)^2 + (\hbar\omega_0)^2} + k_BT] \qquad (31)$$

which is a good approximation of the exact result $\sigma_p^2 = (m\hbar\omega_0/2)\coth(\hbar\omega_0/2k_BT)$. On the plot below the ratio between the approximate and exact expressions is plotted as a function of the dimensionless temperature $\theta = k_BT/\hbar\omega_0$:

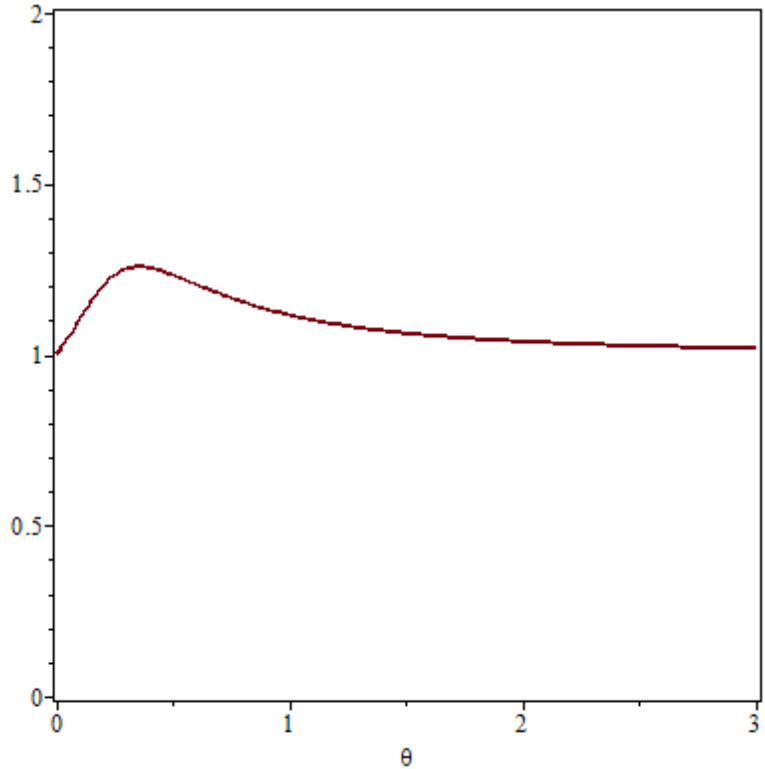

[1]  Bohm, D.: Phys. Rev. **85**, 166 (1952)
[2]  Bohm, D., Vigier, J.P.: Phys. Rev. **96**, 208 (1954)
[3]  Caldirola, P.: Nuovo Cim. **18**, 393 (1941)
[4]  Coffey, W.T., Kalmykov, Y.P., Titov, S.V., Mulligan, B.P.: J. Phys. A Math. Theor. **40**, F91 (2007)
[5]  Davidson, M.: Physica A **96**, 465 (1979)
[6]  Dodonov, V.V.: J. Korean Phys. Soc. **26**, 111 (1993)
[7]  Fenyes, I.: Z. Phys. **132**, 81 (1952)
[8]  Furutsu, K.: J. Res. Natl. Bur. Stand. D **67**, 303 (1963)
[9]  Kanai, E.: Prog. Theor. Phys. **3**, 440 (1948)
[10] Kostin, M.D.: J. Chem. Phys. **57**, 3589 (1972)
[11] Madelung, E.: Z. Phys. **40**, 322 (1927)
[12] Nelson, E.: Phys. Rev. **150**, 1079 (1966)
[13] Neumaier, A.: arXiv: quant-ph/0001011 (2000)
[14] Novikov, E.A.: Sov. Phys.-JETP **20**, 1290 (1965)
[15] Petrosky, T., Prigogine, I.: Chaos Solitons Fract. **4**, 311 (1993)
[16] Petrosky, T., Prigogine, I.: Chaos Solitons Fract. **7**, 441 (1996)
[17] Reginatto, M.: Phys. Rev. A **58**, 1775 (1998)
[18] Sonego, S.: Found. Phys. **21**, 1135 (1991)
[19] Spiller, T.P., Spencer, P.S., Clark, T.D., Ralph, J.F., Prace, H., Prance, R.J., Clippingdale, A.: Found. Phys. Lett. **4**, 507 (1991)
[20] Tsekov, R.: Int. J. Theor. Phys. **48**, 85 (2009)
[21] Tsekov, R.: Int. J. Theor. Phys. **48**, 630 (2009)
[22] Tsekov, R.: Int. J. Theor. Phys. **48**, 1431 (2009)
[23] Tsekov, R.: Int. J. Theor. Phys. **48**, 2660 (2009)
[24] Tsekov, R.: arXiv 0904.0723 (2009)
[25] Turski, L.A.: Acta Phys. Pol. B **26**, 1311 (1995)
[26] Turski, L.A.: Lect. Notes Phys. **477**, 347 (1997)
[27] Wigner, E.: Phys. Rev. **40**, 749 (1932)
[28] Brown, M.R.: arXiv:quant-ph/9703007 (1997)